\documentclass{elsart}
\usepackage{natbib}
\usepackage{graphicx}
\usepackage{amsmath, amssymb}

\begin{document}

\begin{frontmatter}


\title{Power-law scaling in dimension-to-biomass relationship of fish
 schools}

\author{Hiro-Sato Niwa\thanksref{phone}}
\thanks[phone]{Tel.: +81-479-44-5953; fax: +81-479-44-1875.\\
\hspace*{3mm} {\it E-mail address:} Hiro.S.Niwa@fra.affrc.go.jp (H.-S. Niwa).}

\address{Behavioral Ecology Section,
National Research Institute of Fisheries Engineering,
Hasaki, Ibaraki 314-0421, Japan}

\begin{abstract}
Motivated by the finding that there is some biological universality in
 the relationship between school geometry and school biomass of various
 pelagic fishes in various conditions,
 I here establish a scaling law for school dimensions:
 the school diameter increases as a power-law function of school biomass.
The power-law exponent is extracted through the data collapse, and is
 close to $3/5$.
This value of the exponent implies that the mean packing density
 decreases as the school biomass increases, and
the packing structure displays a mass-fractal dimension of $5/3$.
By exploiting an analogy between school geometry and polymer chain
 statistics,
I examine the behavioral algorithm governing the swollen conformation of
 large-sized schools of pelagics, and
 I explain the value of the exponent.
\end{abstract}


\end{frontmatter}

%

\section{Introduction}
\subsection{Brief history of studies on the packing geometry of fish schools}

Animal packing in social aggregations is of fundamental interest in
ecology, and their conformations have been extensively studied
\citep{Parr27,Symons71,Okubo-Chiang74,Graves76,Pitcher-Partridge79,Aoki80,Partridge80,Partridge82,Dill81,Aoki-Inagaki88,misund93a,Parrish-Hamner97}.
Pioneer tank observations of pelagic fishes (herring, sprat and mackerel)
were conducted by \citet{Parr27}, and the ``persistently uniform
density'' of a school was noted.
From experiments on schools of saithe {\it Pollachius virens}, herring
{\it Clupea harengus} and cod {\it Gadus morhua} cruising in a ten
meters circular gantry tank,
\citet{Pitcher-Partridge79} validated that all the fish in a
school occupy a volume proportional to $Nb^3$,
where $N$ is the number of fish and $b$ is the mean nearest-neighbor
distance (approximate to the mean fish-body length) in the school.
\citet{misund93a} reported from field observations that 
the number densities of herring schools are almost independent of the
dimensional size of school
but are an order of magnitude lower than the packing densities observed
when schooling in small tanks.

Tauti and colleagues
\citep{Tauti-Miyosi29,Tauti-Hudino29,Tauti-Yasuda29,Tauti-Yasuda30,Tauti-Yasuda33a,Tauti-Yasuda33b}
had experimentally shown that a fish school can be viewed as a group of
inorganic particles and treated with the methods of physics.
\citet{breder54} explicitly applied physical equations to such
fish schools.
Recently,
a number of theoretical and numerical models of schooling and flocking
behavior have been studied by biologists, mathematicians and physicists
\citep{okubo-levin01,camazine-etal01,vicsek01-book}.
As for the animal group geometry,
many models for social aggregations, however, predict that densities
increase as the group size (in number) increases
[overviews of such models are given in
Refs. \citep{Warburton-Lazarus91,Beecham-Farnsworth99}].
\citet{Mogilner-etal03} mathematically tackled this problem of constant
density and revealed the condition for a well-spaced group, i.e.
what class of mutual interactions results in a relatively constant
individual distance in the interior of the aggregate.

More recently, by means of underwater acoustics,
the school sizes (in number or biomass of fish) have been quantitatively
measured for different values of the dimensional size of schools in the
wild.
Precise data on conformations of large-sized schools of pelegic fishes
became available rapidly,
which were extremely helpful in elucidating a certain geometric law, i.e.
the relation between dimensional and biomass sizes of pelagic fish
schools,
bringing about some important changes in our viewpoints.
Misund and
colleagues \citep{Misund90,Misund-etal92,Misund93b,Misund-etal95,Misund-etal96,Coetzee00,Misund-Coetzee00,Misund-etal03}
found that the power-law scaling in dimension-to-biomass
relationship exists, and is robust across a broad range of pelagic species
as well as across diverse environments.
They demonstrated that 
if the biomass $N$ in a school is, say, doubled,
the cross-sectional area of a school is increased by a
factor $2^{2\nu}$, i.e.
\begin{equation}
 \mbox{cross-sectional area} \propto (\mbox{biomass})^{2\nu},
  \label{eqn:misund-school-geometry}
\end{equation}
and that the exponent $\nu$ looks universal,
reading $0.5$
(from the field data $\nu$ ranges from $0.415$ to $0.77$).
The geometric law they found implies that
the mean density of a school scales as $N^{1-3\nu}$ in three dimensions
of space and the conformation of social aggregations swells
(i.e. $\nu > 1/3$).
Such a relationship as Eq.(\ref{eqn:misund-school-geometry}) with
$\nu = 0.5$ has been utilized for the dimensions to biomass conversion
(e.g. transforming the school diameter $R$ to the number $N$ of fish)
in analyzing school-size distributions
\citep{Anderson81,Niwa96b,Niwa98,Niwa03,Niwa04a}.

It might come as a surprise that
packing densities decrease as the group size increases,
contrary to previous observations and predictions.
Laboratory observations for school geometry and internal structure have
been made exclusively in small tanks,
which generally show the constant density of fish that the school
volume is proportional to the number of individuals.
In all set-ups, the factors resulting in homogeneous, cohesive school
may be especially pronounced, and
it therefore seems that the quantified structure is skewed.
\citet{Pitcher-Parrish93} claimed that homogeneity in schools has been
over-emphasized.
{\it In-situ} observations of herring and sprat schools with a high
resolution sonar revealed that the packing structure within the schools
is rather heterogeneous \citep{Cushing77}.
This has been confirmed by measurements of free-swimming schools using
photography and high-resolution echo integration, which showed that the
packing density distribution in capelin {\it Mallotus villosus} and
clupeoid schools varies considerably \citep{Freon-etal92,misund93a}.
Regions of high density are usually found within the schools, and even
empty vacuoles have been recorded.
\citet{Misund-Floen93} observed by repeated echo integration that there
were large variations in internal packing density of herring schools
(i.e. high-density regions or empty lacunas within a school),
and that the packing density structure was quasi-stationary.
Besides artificial environments in small tanks,
the discrepancy in former observations could have been caused by too
small numbers of fish in the schools.
For instance, in \citet{Pitcher-Partridge79},
$N$ takes a few tens of fish.
Since such a geometric law above is always defined only in a certain
limit \citep{deGennes79},
the scaling in the dimension-to-biomass relationship is expected to hold
for large-sized schools of pelagics.

In this paper, the exponent $\nu$ is estimated according to
the established universal scaling law in the school-size distribution
of pelagic fishes \citep{Niwa03,Niwa04a}:
choose the suitable value of $\nu$ to achieve the best data-collapse on
the size distributions in terms of the school dimension.
Notice that the dimension-to-biomass relationship is a property of the
single school and the scaling exponent $\nu$ is determined by the
behavioral algorithm of fish schooling at individual level,
while the scaling in the school-size distribution emerges from the
inter-school interactions at population level
(i.e. a global property of the interacting school system).

\subsection{Scaling in school-size distributions}

\begin{figure}[tb]
 \centering
 \includegraphics[width=6.5cm]{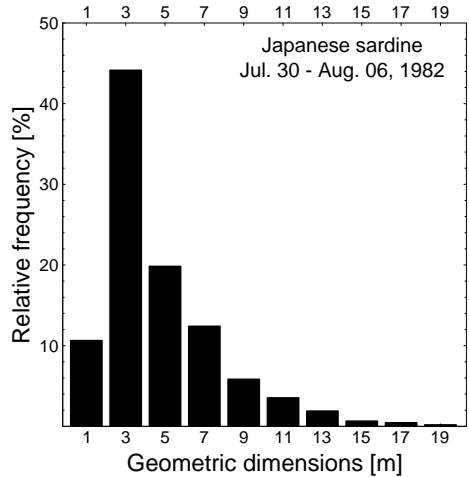}
 \caption{
 Vertical dimension (in meters) of Japanese sardine schools.
 Data from the acoustic survey by \citet{Hara90} off southeastern
 Hokkaido in the summer, 1982.
 The same data are available in \citet{Hara84}.
 Small schools of vertical dimensions in the first bin
 ($<\mbox{two meters}$) may be stray ones.
 }
 \label{fig:1}
\end{figure}

Animal group size is a focal issue in ecology that,
in contrast to scaling,
has introduced a single preferred size
(i.e. optimal or compromise size)
for any organism living in groups
\citep{Pulliam-Caraco84,Higashi-Yamamura93,Niwa96b,Hoare-etal04}.
Figure~\ref{fig:1} shows an example of the histogram of school
dimensions of Japanese sardine {\it Sardinops melanostictus}.
A peak frequency and a right skew are typical of pelagic fishes
\citep{Anderson81,Niwa96b,Niwa98}.
Their linear dimension, e.g. the vertical thickness, of the school
falls into a certain range below a few tens of meters.
One fish may not be the right atomic unit in schooling, since field
observations suggest that no school exists under a certain minimal size.
This may cause binned data of school sizes to exhibit the fake peak
frequency.
\citet{Anderson81} and \citet{Niwa96b} have ignored
slowly decaying (or fat-tailed) distributions, including scaling laws,
in such data of school sizes short ranged with a fake peak frequency
(the possibility of power laws was already presented in their models but
not exploited).
A traditional, widely used Gauss statistics says that, for the data from
\citet{Hara90} shown in Fig.\ref{fig:1},
finding sardine schools ranging from 18 to 20 meters in vertical
thickness should only occur about once every $10^9$ detections of
schools
[for detail, consult \citet{Niwa04a}].
In other words, it is not the real world!
Aquatic observations actually say that finding such schools occurs about
once every 500 detections.
The probability that such schools are found is $10^6$ times as large!
%

The possibility of scaling in such distributions was found by
\citet{Bonabeau-Dagorn95}.
Fat-tailed distributions have been found to quite generally characterize
the size heterogeneity of pelagic fish schools
\citep{Niwa98,Bonabeau-etal98,Bonabeau-etal99}.
Only lately, \citet{Niwa03,Niwa04a} showed that the school-size
distribution of pelagic fishes has the universal functional form of a
power-law decay and a crossover towards the exponential distribution.

\subsection{Scaling and data collapse in ecology}

Scaling laws have been found to characterize many patterns
in ecological systems \citep{Azovsky00,Chave-Levin03,Halley-etal04}.
There is an empirical rule about the relationship between the areas
($L$) of island and the numbers ($S$) of species on them:
whether counting birds, butterflies, plants or copepods in ponds, a
tenfold smaller area contains roughly half the species.
This can be fitted by a power-law function
\begin{equation}
 S(L)\propto L^z
  \label{eqn:species-area-rel}
\end{equation}
with the exponent $z\approx 0.25$
\citep{Preston62,MacArthur-Wilson67,Harte-etal99,May99}.
The species-area relationship is widely used in rough assessments of
likely future rates of species extinction
(because observed rates of tropical deforestation are equated to
loss of habitat area,
we can assess the annual production of species committed to
extinction).
So, a more secure understanding of such relationships has great
practical importance \citep{May99}.
\citet{Banavar-etal99} applied finite-size scaling
\citep{Fisher-Barber72,Binder-Heermann88}
to test the hypothesis of scaling invariance in the species-area
relationship resting on a model proposed by \citet{Harte-etal99} for the
species abundance distribution.
Following the conventional notation in the physics literature, the
system sizes are hereafter denoted by $L$.

The finite-size scaling (FSS) hypothesis assumes that
the fraction of objects with size $n$ for a finite system of size $L$ is
written, with a scaling exponent $A$, as
\begin{equation}
 P(n;L)
  =
  L^{-A} F\left(n/L^A\right),
  \label{eqn:FSS-hypothesis}
\end{equation}
where $F$ is a universal scaling function.
$P(n;L)$ reads, for instance,
the species-abundance distribution, which defines
the probability that any given species on a census patch of area $L$
has $n$ individuals,
where
$P(n;L)S(L)$ is the expected number of species with $n$ individuals.
$L^A$ is a measure of the width of the probability distribution
$P(n;L)$,
e.g. the mean or the standard deviation of the distribution.
Equation~(\ref{eqn:FSS-hypothesis}) expresses the principle that 
the behavior of the finite system, i.e. $P(n;L)$, is controlled by the
ratio $n/L^A$.
We test the FSS hypothesis by data collapsing:
when
$y = L^A P(n;L)$
is plotted versus
$x = n/L^A$,
if the distributions across different $L$'s fall on a single curve
(e.g.
collapsing distinct abundance distributions measured in different
areas and with different total numbers of individuals onto one scaling
curve),
then we should be able to identify a scaling function $F(x)$.
In order to determine the exponent $A$,
we try a best-fitting procedure such that the family of curves
$P(n;L)$ collapses onto a single curve as well as possible.
We then obtain other scaling exponents (e.g. $z$ of the species-area
relationship) resting on the scaling relation between scaling
exponents \citep{Goldenfeld92}.
By noting that
the total number of individuals of all species is equal to
$\sum_n n P(n;L)S(L)$,
the species-area relationship leads to a scaling relation
\begin{equation}
 A+z=1,
  \label{eqn:scaling-ecology}
\end{equation}
assuming that
the total number of individuals of all species is proportional
to area $L$ \citep{Banavar-etal99}.

The universal scaling is fundamental to data collapse.
The power-law scaling emerges as ubiquitous properties in ecology.
In statistical analysis in ecology, data collapsing across different
environments or species has been, however, observed only lately in
dynamics of breeding bird populations
\citep{Keitt-Stanley98,Keitt-etal02},
food web structure \citep{Camacho-etal02},
microbial body-mass spectra in marine ecosystems \citep{Rinaldo-etal02},
school-size distributions of pelagic fishes \citep{Niwa03,Niwa04a},
and in the context of ecological economics,
capture fisheries productions of countries \citep{Niwa04b}.

\section{Scaling in dimension-to-biomass relationship}
\subsection{Estimation of the exponent $\nu$ through data collapse}

In order to understand the geometric properties of school
configurations, the school biomass $N$ must be measured for different
values of dimensional size $R$, and we must compare them.
Misund and colleagues extensively performed the simultaneous
observations of the values of $R$ and $N$ for pelagics by the following
two methods:
(i)~two-dimensional (cross-sectional) acoustic measurements (unit in
square meters) and subsequent purse seine capture (unit in tonnes) of
schools \citep{Misund90,Misund93b},
and
(ii)~acoustic measurements of the three-dimensional structures and
backscattered echo energy \citep{Misund-etal92,Misund-etal95,Misund-etal96,Coetzee00,Misund-Coetzee00,Misund-etal03}.
The surveys were conducted on stocks of
anchovy {\it Engraulis capensis},
herring {\it Clupea harengus},
horse mackerel {\it Trachurus trachurus capensis},
mackerel {\it Scomber scombrus},
pilchard {\it Sardinops ocellatus},
round herring {\it Etrumeus whiteheadi},
saithe {\it Pollachius virens},
sardine {\it Sardinops sagax},
and
sprat {\it Sprattus sprattus},
in different seasons and geographic regions
(the Barents Sea,
the North Sea,
the Norwegian Sea,
the northeastern Atlantic and off Namibia,
and
off the coast of South Africa).
They found that there is some biological universality in
the dimensions-to-biomass relationships,
Eq.(\ref{eqn:misund-school-geometry}).
Their finding implies that
the radius $R$ of the school scales as
\begin{equation}
 R = (\mbox{constant})\times N^{\nu}
  \label{eqn:scaling-school-geometry}
\end{equation}
in a statistical sense.
$N$, denoting the school biomass, is hereafter defined by the number of
fish in a school.

Now I propose another way to establish the scaling law for school
dimensions, Eq.(\ref{eqn:scaling-school-geometry}), by applying
universal scaling law in fish school-biomass distributions
\citep{Niwa03,Niwa04a}.
The biomass distributions $W(N)$ follow a power law with exponent
$\beta =1$
up to a cut-off size
$\langle N \rangle_P$,
\begin{equation}
 W(N)
  =
  N^{-\beta} P \left(N/\langle N\rangle_P\right),
  \label{eqn:biomass-distribution}
\end{equation}
where $P(x)$ is a crossover scaling function with a strong drop for
$x>1$, and the cut-off size
(crossover size from power-law to exponential decay)
is calculated from the biomass histogram data
 $\{(N_i,W_i)|i=1,2,\ldots\}$,
\begin{equation}
 \langle N\rangle_P
  =
   \frac{
   \sum_i N_i^2 W_i\Delta N
   }{
   \sum_i N_i W_i\Delta N
   },
   \label{eqn:cut-off-size}
\end{equation}
where histogram bins are chosen with width $\Delta N$. 

The dimension data of fish schools are binned with width $\Delta R$,
giving the set of frequencies
$\left\{\left.\left(R_i,W^{\mbox{\tiny (G)}}_i\right)\right| i=1,2,\ldots\right\}$.
From Eqs.(\ref{eqn:scaling-school-geometry}) and
(\ref{eqn:biomass-distribution}),
the distribution of geometric dimensions of fish schools is represented
as
\begin{equation}
 W^{\mbox{\tiny (G)}}(R)
  = 
  R^{-1}
  P^{\mbox{\tiny (G)}}\left(R/{\langle R\rangle_P}\right),
  \label{eqn:dimension-distribution}
\end{equation}
where
\begin{equation}
 \langle R \rangle_P
  =
  \langle N \rangle_P^{\nu},
\end{equation}
and
\begin{equation}
 P^{\mbox{\tiny (G)}}(x)
  =
  P(x^{1/\nu}).
\end{equation}
Therefore, the school-dimension distribution follows a power law with
the same exponent ``$-1$'' as the school-biomass distribution.
The power-law distribution $W^{\mbox{\tiny (G)}}(R)$ is truncated at a
cut-off size $\langle R \rangle_P$,
which is also calculated from histogram data of school geometric
dimensions,
\begin{equation}
 \langle R\rangle_P
  =
  \left[
   \frac{
   \sum_i R_i^{2/\nu} W^{(\mbox{\tiny G})}_i \Delta R
   }{
   \sum_i R_i^{1/\nu} W^{(\mbox{\tiny G})}_i \Delta R
   }
 \right]^{\nu}.
  \label{eqn:depend-cutoff-nu}
\end{equation}
The following normalizations are adopted for the scaling function
$P^{\mbox{\tiny (G)}}(x)$ and
the histogram data of geometric dimensions of fish schools, because the
cut-off size $\langle N \rangle_P$ is proportional to the total number
of fish in the school system
[\citet{Niwa03,Niwa04a}; see also Eq.(\ref{eqn:cutoff-population})]:
\begin{equation}
\int_0^{\infty} x^{1/\nu-1} P^{\mbox{\tiny (G)}}(x)\mbox{d}x
 = 1,
 \label{eqn:dimension-distribution-normalization}
\end{equation}
and
\begin{equation}
 \sum_i R_i^{1/\nu} W^{\mbox{\tiny (G)}}_i \Delta R
  =
  \langle R\rangle_P^{1/\nu},
\end{equation}
respectively.

\begin{figure}[tb]
 \centering
 \includegraphics[width=6.5cm]{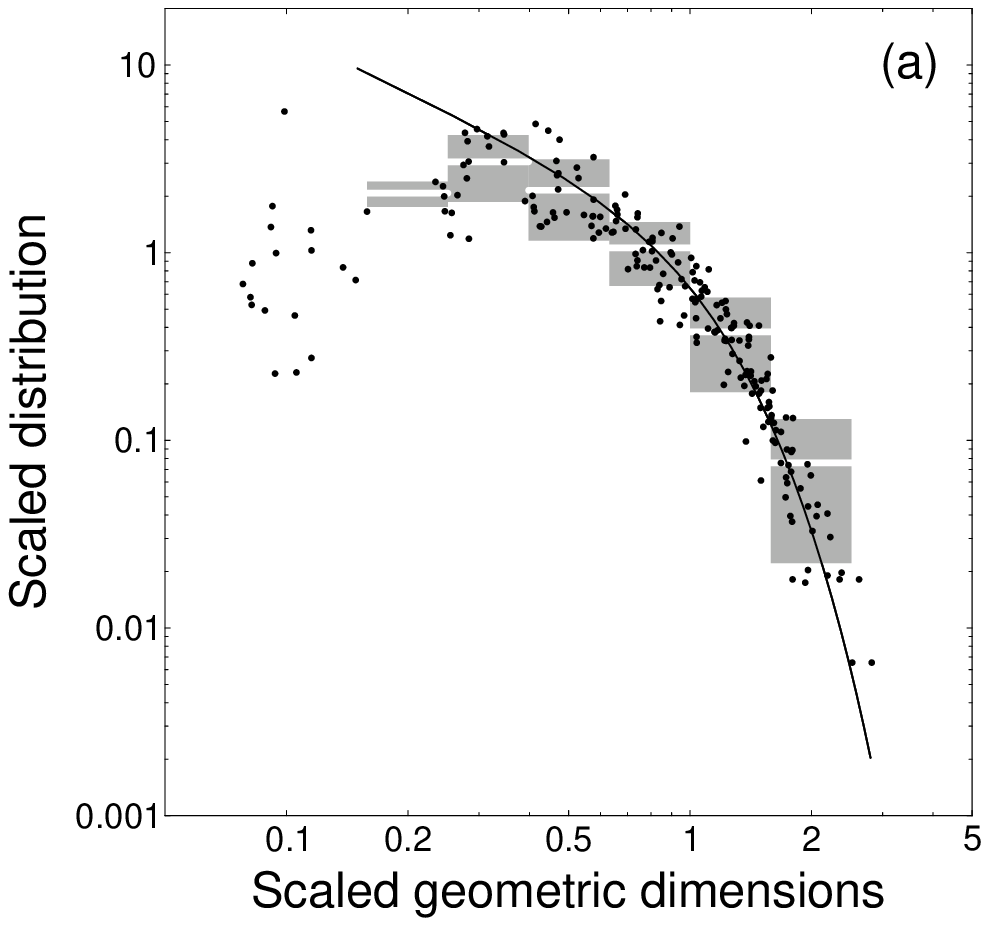}
 \hspace{0.5cm}
 \includegraphics[width=6.5cm]{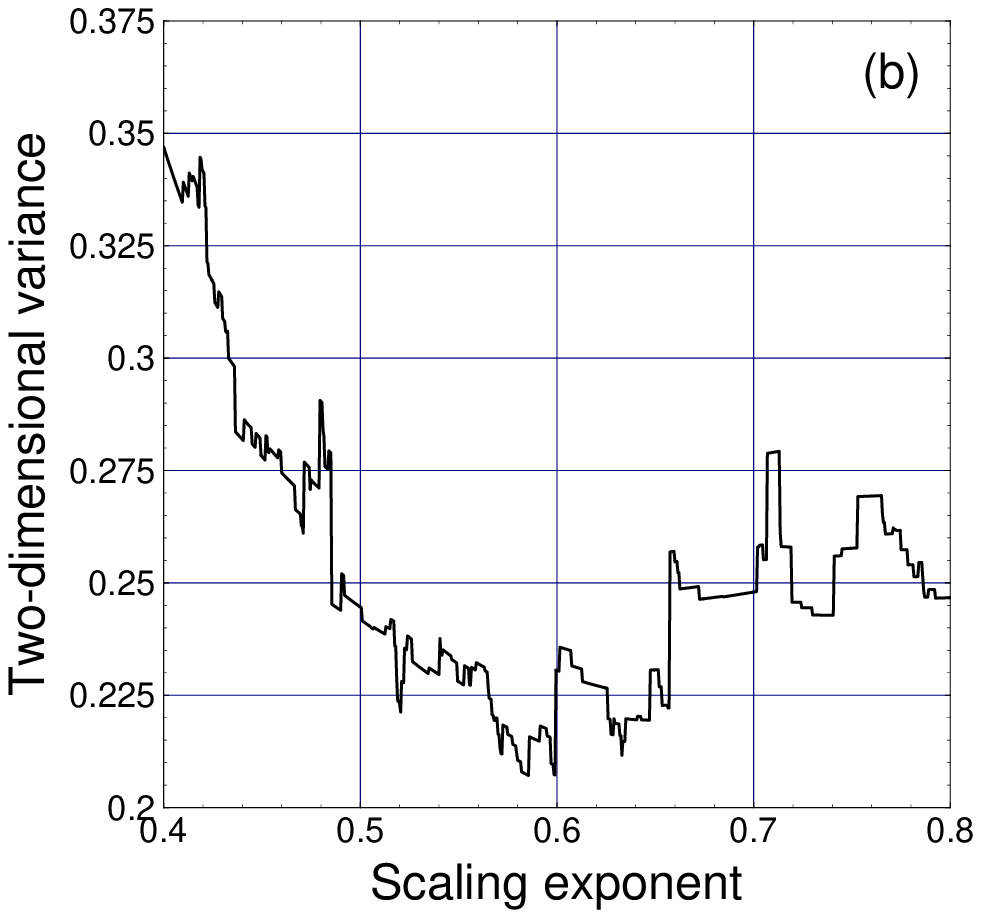}
 \caption{
 Data collapse to extract the exponent $\nu$.
 (a)~Scaled distribution of geometric dimensions of sardine schools.
 $y= W^{\mbox{\tiny (G)}}\langle R\rangle_P$
 is plotted versus
 $x= R/\langle R\rangle_P$
 with $\nu = 0.6$ on double-logarithmic scale.
 The bins are chosen equally spaced on a logarithmic scale as
 $x \in \left[10^{-1+k/5}, 10^{-1+(k+1)/5}\right)$ with
 $k=1,2,\ldots,6$.
 The rectangle in gray reads the interval $\mu_y\pm\sigma_y$,
 i.e. the error $\sigma_y$ on the mean $\mu_y$ (indicated by the slit)
 for each bin.
 The solid line is a prediction of the mean-field theory \citep{Niwa03}.
 (b)~The mean of two-dimensional variance,
 $\overline{\epsilon}$,
 versus the power-law exponent
 $\nu$.
 Although $\overline{\epsilon}$ shows noisy fluctuations, it takes a
 minimum around $\nu= 0.6$.
 Data from \citet{Hara90} are analyzed.
}
 \label{fig:2}
\end{figure}

Fat-tailed school-size distributions are necessarily truncated because
the population is finite.
Since the size $\langle R\rangle_P$ depends on the exponent $\nu$
following Eq.(\ref{eqn:depend-cutoff-nu}), so that we can determine the
value of $\nu$ by evaluating the location of the cut-off in the
power-law distribution $W^{\mbox{\tiny (G)}}(R)$.
The ordinary least squares regressions might, however, lead to a
``wrong'' estimation of the exponent
\citep{Niwa98,Bonabeau-etal98,Bonabeau-etal99}.
I make use of the data collapse to extract the ``right'' exponent.
From Eqs.(\ref{eqn:dimension-distribution}) and
(\ref{eqn:dimension-distribution-normalization}), when
$
y=W^{\mbox{\tiny (G)}}\langle R\rangle_P
$
is plotted against
$x=R/\langle R\rangle_P$
with correct parameter $\nu$,
all the empirical data should collapse onto each other.
The power-law exponent of school-dimension distributions, $\nu$, is
then evaluated through data collapse.
Let us search for the value  of $\nu$ that places all the points most
accurately on a single curve.
We use a set of histogram data of vertical dimension of Japanese sardine
{\it Sardinops melanostictus} schools, from 22 acoustic surveys by
\citet{Hara90} off southeastern Hokkaido for seven years, 1981--1987.
\citet{Hara86} reported that
Japanese sardine migrate as a huge-sized school in number from a few
hundreds of thousands to a few million of fish.
To obtain the best data collapse,
the $x$-axis is divided into bins (Fig.\ref{fig:2}a), and
for each bin two-dimensional variance
\begin{equation}
 \epsilon
  =
  \left(\sigma_x/\mu_x\right)^2
  +
  \left(\sigma_y/\mu_y\right)^2
  \label{eqn:2D-var}
\end{equation}
is calculated, where $\sigma_x$ and $\sigma_y$ denote the standard
deviation of the mean $\mu_x$ and $\mu_y$, respectively.
The parameter $\nu$ is then estimated at value that minimize the mean of
two-dimensional variance for the bins (Fig.\ref{fig:2}b).
The mean of two-dimensional variance,
$\overline{\epsilon}$, is a measure to determine the goodness of
collapse \citep{Lillo-etal02,Lillo-etal03,Niwa04a}.
A good data collapse can be obtained by using the value
$\nu \approx 0.6$.
The resulting plot of empirical school data is shown in
Fig.\ref{fig:2}a.
Experimentally fitting the parameter $\nu$ to achieve a good data
collapse, ``$3/5$'', is reminiscent of the Flory value of the
exponent in a power-law dependence of the coil radius of a polymer chain
(in three-dimensional solutions) on the degree of
polymerization \citep{Flory53,deGennes79}.
%

Here
we see that the power-law regime of the distribution is too short,
which misled \citet{Anderson81} and \citet{Niwa96b} into overlooking the
power-law distributions of school sizes.
Notice that many power laws that appeared in the ecology literature
span less than two orders of magnitude of scale \citep{Halley-etal04}.
The power law range of too few scales is not unique to ecology;
the largest numbers of power laws reported in the physical science are
for small ranges \citep{Hamburger-etal96}.

\subsection{Retest of the FSS in school-biomass distributions}

\begin{figure}[tb]
 \centering
 \includegraphics[width=6.5cm]{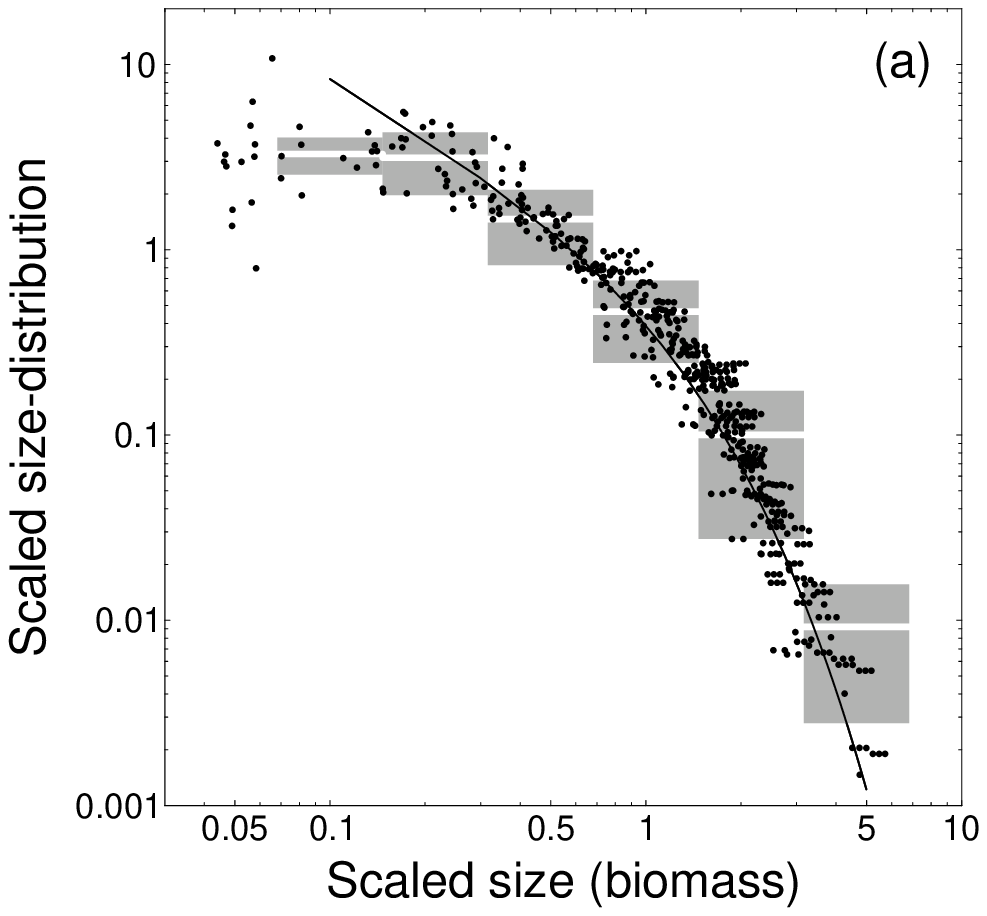}
\hspace{0.5cm}
 \includegraphics[width=6.5cm]{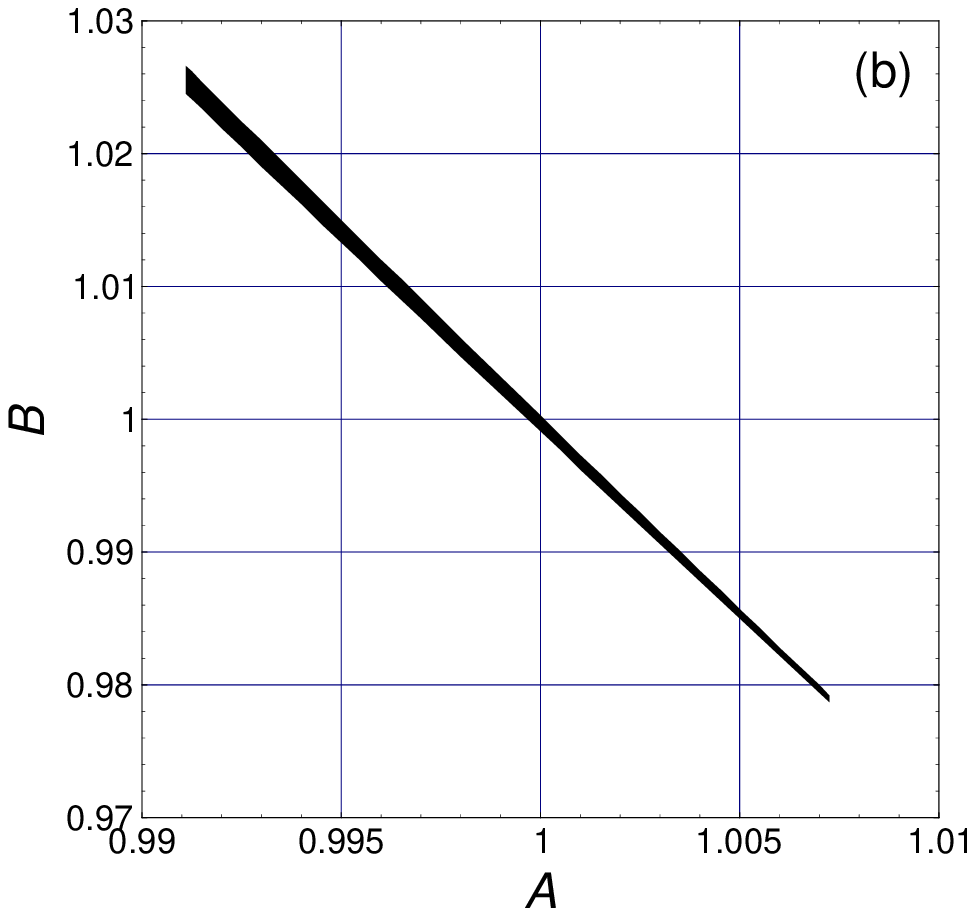}
 \caption{
 FSS analysis of school-size data.
 (a)~FSS plot of the biomass distribution on double-logarithmic
 scale.
 Dimension data of sardine schools (same as Fig.\ref{fig:2}) are converted
 by Eq.(\ref{eqn:biomass-dimension-convert}) with $\nu =3/5$.
 Here $y= W L^{A + B}$
 is plotted versus
 $x= N/L^A$
 with $A =B =1$.
 Two-dimensional variances [same as Eq.(\ref{eqn:2D-var})] are
 calculated for bins chosen equally spaced on a logarithmic scale as
 $x \in \left[10^{-1+(k-0.5)/3}, 10^{-1+(k+0.5)/3}\right)$
 with $k=0,1,\ldots,5$.
 The rectangle in gray is same as Fig.\ref{fig:2}.
 The solid line is a prediction of the mean-field theory \citep{Niwa03}.
 (b)~The region of the $AB$-plane in which
 the minimum of the mean of two-dimensional variance exists.
 A measure of data collapse for scaling, i.e. the mean of
 two-dimensional variance, $\overline{\epsilon}$, takes a minimum
 $\overline{\epsilon}_{\mbox{\scriptsize min}}$ for the right choice of
 $(A, B)$.
 The minimum is found with the precision i.e. width of the minimum,
 $\Delta\epsilon = 10^{-3}$ in black region
 ($
 \Delta\epsilon/\overline{\epsilon}_{\mbox{\scriptsize min}}
 \approx 3.12\times 10^{-3}
 $).
 The values of the parameters lie in the intervals
 $A = 0.999\pm 0.008$ and $B = 1.003\pm 0.024$,
 and therefore
 $\beta = 1.004\pm 0.032$.
}
 \label{fig:3}
\end{figure}

The acoustic-survey data are converted into a school-biomass histogram
as follows
\begin{equation}
 W(N) \Delta N
  \propto
  W^{\mbox{\tiny (G)}}(R) R^{1/\nu -1} \Delta R.
  \label{eqn:biomass-dimension-convert}
\end{equation}
We now crosscheck the value of $\nu$ through finite-size scaling (FSS)
analysis of school-biomass distribution \citep{Niwa04a}.
Since the finite population size causes the truncation of power-law
distribution
$W(N) \propto N^{-\beta}$,
there is a well-defined quantity
\begin{equation}
 L
  =
  \frac{
  \sum_i N_i^{1+\beta} W_i\Delta N
  }{
  \sum_i N_i^{\beta} W_i\Delta N
  },
  \label{eqn:sys-size-school}
\end{equation}
which depends on the system population size.
In order to characterize the finite size effects,
FSS hypothesis is used:
the distribution function depends on $N$ only through the ratio $N/L^A$,
\begin{equation}
 W(N; L) \mbox{d}N
  =
  L^{-B}
  F \left(N/L^A\right) \mbox{d}\left(N/L^A\right),
\end{equation}
where $F(x)$ is a universal function independent of fish population size.
The prefactor $L^{-B}$ is required to ensure the normalization
\begin{equation}
 \sum_i N_i^{\beta} W_i\Delta N = 1,
\end{equation}
where $P(N)\;[\equiv N^{\beta} W(N)]$ now represents the probability
distribution of school-biomass sizes.
From the FSS hypothesis, it is expected that when
$W L^{A+B}$
is plotted against
$N/L^A$
with correct parameters $A$ and $B$
all the data collapse onto a single curve.
The power-law exponent of biomass distributions, $\beta$, is
then evaluated through FSS analysis.
The value of $B/A$ is the estimate of the power-law exponent
\begin{equation}
 \beta = \frac BA.
\end{equation}
Let us analyze a set of 22 acoustic-survey data of sardine schools (same
as Fig.\ref{fig:2}) converted into biomass histograms by using
Eq.(\ref{eqn:biomass-dimension-convert}) with $\nu =3/5$.
In a simultaneous best-fitting procedure (Fig.\ref{fig:3}),
a good data collapse can be obtained by using the values
$A \approx 1$ and $B \approx 1$.
The power-law exponent derived from the FSS collapse is $\beta \approx 1$. 
The resulting plot is shown in Fig.\ref{fig:3}a.
The school-biomass distribution follows a power-law decay with exponent
$-1$, and is truncated at the cut-off size of Eq.(\ref{eqn:cut-off-size}).
The FSS collapse confirms the scaling laws for school sizes,
Eq.(\ref{eqn:scaling-school-geometry}) with $\nu =3/5$ and
Eq.(\ref{eqn:biomass-distribution}) with $\beta =1$.
%

Finite-size scaling techniques have been applied to ecology and shown to
predict scaling relations between scaling exponents
in relative abundance of species [Eq.(\ref{eqn:scaling-ecology})]
\citep{Banavar-etal99,Aji-Goldenfeld01},
in dynamics of breeding bird populations \citep{Keitt-etal02},
in biomass-size distributions of seston \citep{Rinaldo-etal02},
and in exploitation of fish stocks \citep{Niwa04b}.
In school-size statistics, we expect to have a scaling relation.
We now choose the normalization
\begin{equation}
 \sum_i N_i W_i\Delta N = \Phi,
  \label{eqn:norm-phi}
\end{equation}
where $\Phi$ denotes the total fish population in the school system
($\sum_i W_i\Delta N$ gives the total number of schools).
Since Eq.(\ref{eqn:sys-size-school}) implies 
$L^A \propto \Phi^{\gamma}$ with a scaling exponent $\gamma$,
the FSS relation for the school-biomass distribution is written as
\begin{equation}
 W(N)
  =
  N^{-\beta} P(N/\Phi^{\gamma}).
\end{equation}
Accordingly, the normalization of Eq.(\ref{eqn:norm-phi}) yields the
scaling relation
\begin{equation}
 2 -\beta =1/\gamma.
\end{equation}
The best-fitting value in data collapse, $\beta\approx 1$, gives
$\gamma\approx 1$.
As a consequence,
the location of the cut-off in the power-law distribution of school
sizes simply reads
\begin{equation}
 \langle N \rangle_P
  \propto
  \Phi,
  \label{eqn:cutoff-population}
\end{equation}
which is verified by means of extensive numerical simulations
\citep{Niwa03,Niwa04a}.

\section{Behavioral algorithm of fish schooling}
\subsection{Gaussian model}

Let us now investigate cohesive motion of schools in a three-dimensional
space from the viewpoint of the behavioral algorithms which govern their
formation and dynamics:
attraction between neighbors maintains cohesion of the school;
a tendency to align with neighbors produces collective motion of the
school.
The minimal model of cohesion is a linkage of neighbors consisting of
harmonic spring, because the inter-fish distance follows a Gaussian
distribution \citep{Dill81,Partridge82,Niwa96a}.
Using the relative coordinates to the center of school,
the equation of motion of fish (as noisy self-propelled particles) in a
large school of size $N$ is written by a one-body approximation as the
following Langevin equation \citep{Niwa94,Niwa96a}:
\begin{equation}
 \frac{\mbox{d}^2\boldsymbol{r}}{\mbox{d}t^2}
  =
  \boldsymbol{f} (\boldsymbol{r})
  -J\frac{\mbox{d}\boldsymbol{r}}{\mbox{d}t}
  +\boldsymbol{\eta}(t),
  \label{eqn:langevin}
\end{equation}
providing that individuals are sufficiently sensitive to behavior of
their neighbors,
where $\boldsymbol{f}(\boldsymbol{r})$ is the cohesive force;
$J$ denotes the strength of alignment;
and
$\boldsymbol{\eta}$ is a random perturbation of the velocity with
strength $\epsilon_v$ and a $\delta$-correlated time dependence.

\subsection{Self-consistent calculation of of cohesive force}

The attractive interaction force acting on one body due to a system of
bodies is the neighbor-joining harmonic spring.  One essential
approximation is to replace the many-fish problem by the problem of
solving the motion of one fish in a certain self-consistent field.  The
cohesive force is then written in the following form by the one-body
approximation:
\begin{equation}
 \boldsymbol{f} (\boldsymbol{r})
  =
  -\frac{\tilde{k}}{N^{\alpha}} \boldsymbol{r},
  \label{eqn:cohesion}
\end{equation}
where $\tilde{k} N^{-\alpha}$ is the effective spring constant for the
springs strung out from the center of the school to a fish:
the number of consecutive springs joining the fish to the center
of the school via other companions is proportional to $N^{\alpha}$.
The total ``elastic energy'' of inter-fish bonds in the school depends
linearly on number $N$ of individuals in the school.
The overall elastic energy after integration over a sphere of radius $R$
(denoting the average radius of the school) results in
$
 E_{\mbox{\scriptsize el}}
  \propto
  {R^4}{N^{-\alpha}}
$,
which is derived from the ideal harmonic spring, Eq.(\ref{eqn:cohesion}).
Therefore, the exponent $\alpha$ will be determined self-consistently,
\begin{equation}
 \frac{R^4}{N^{\alpha}}
  \propto
  N.
  \label{eqn:self-consistent-spring}
\end{equation}
The solution of the Langevin equation~(\ref{eqn:langevin}) with
Eq.(\ref{eqn:cohesion}) takes the following asymptotic forms
\citep{hori77}:
\begin{equation}
 \sigma_v^2\,
  \left[\,
   \equiv
   \left\langle\left(
		\mbox{d}\boldsymbol{r}/\mbox{d}t\right
		)^2\right\rangle\right]
   \approx
   {3\epsilon_v}/J,
  \label{eqn:fluctuation-dissipation}
\end{equation}
\begin{equation}
 \langle\boldsymbol{r}^2\rangle
  \approx
  \frac{\sigma_v^2}{\tilde{k}}N^{\alpha},
  \label{eqn:mean-square-radius}
\end{equation}
where the root-mean-square
$\sqrt{\langle \boldsymbol{r}^2 \rangle}$
gives the average radius of the school, $R$.
From Eq.(\ref{eqn:self-consistent-spring}) together with
Eq.(\ref{eqn:mean-square-radius}),
the self-consistent value of the exponent $\alpha$ is obtained:
\begin{equation}
\alpha = 1.
\end{equation}
As a consequence, the self-consistent cohesive force reads
\begin{equation}
 \boldsymbol{f} (\boldsymbol{r})
  =
  -\frac{\sigma_v^2}{b^2 N}\boldsymbol{r},
\end{equation}
where $b$ denotes the effective inter-fish distance
(a constant independent of $N$):
$b^2\equiv \sigma_v^2/\tilde{k}$.
The average radius of the school is then given by
$bN^{1/2}$.
Note that
Eq.(\ref{eqn:fluctuation-dissipation}) is an example of a more general
principle called the fluctuation-dissipation theorem \citep{kubo66}.

Notice that the above model has been developed in the absence of the
excluded volume interactions.
Social cohesion by harmonic spring interactions between neighboring
fish reminds us of the classical picture of a polymer chain based on 
the bead-spring model \citep{Rouse53}.
The exponent $\nu$ close to the value $3/5$ may then be understood by
taking the excluded volume effect \citep{Doi-Edwards86}.
Such a school without excluded volume effect may be called the
``Gaussian'' school in line with polymer physics.

\subsection{The excluded volume effect}

From Eq.(\ref{eqn:langevin}),
the probability of the school radius being between $r$ and $r+\mbox{d}r$ is
given by the following \citep{hori77}:
\begin{equation}
 \Psi_0 (r)
  =
  4\pi r^2
  \left(\frac 3{2\pi b^2N}\right)
  \exp\left(
       -\frac{3r^2}{2b^2N}
     \right)
  \label{eqn:prob-gaussian-radius}
\end{equation}
in a stationary state in three dimensions
(i.e. the position vector $\boldsymbol{r}$ follows a Gaussian
distribution).
The Gaussian school model considered above permits fish to occupy the
same region in space.
Of course this is a physical impossibility since each fish possesses its
own finite volume.
Therefore, in the ``excluded volume'' school, there are a number of
Gaussian school configurations which are disallowed due to the steric
effect.
Let $p(r)$ be the probability that a Gaussian school configuration, as
counted in Eq.(\ref{eqn:prob-gaussian-radius}), is also allowable under
the excluded volume condition.
We now calculate the probability that no overlaps occur when we place
$N$ fish within a region of volume ($\sim r^3$), which will lead to an
estimation for $p(r)$.
The approach is due to \citet{Doi96}.
Letting $w$ be the volume which is effectively excluded to one fish by the
presence of another ($w \lesssim b^3$), the probability that one
particular fish will not overlap with another is then given by
$(1-w/r^3)$.
Since there are $N(N-1)/2$ possible combinations of pairs, the
probability that no overlap occurs in all of these combinations is given
by
\begin{equation}
 p(r)
  =
  (1-w/r^3)^{N(N-1)/2}
  =
  \exp \left(
	-\frac{wN^2}{2r^3}
       \right),
  \label{eqn:prob-no-overlap-1}
\end{equation}
where $r^3 \gg w$ and $N\gg 1$.
Therefore, the probability distribution of the school radius $r$ can
then be written as
\begin{equation}
 \Psi (r)
  =
  \Psi_0(r) p(r)
  \propto
  r^2
  \exp\left(
       -\frac{3r^2}{2b^2N}
       -\frac{wN^2}{2r^3}
      \right)
  \label{eqn:prob-excluded-radius}
\end{equation}
for the excluded volume school.

Both $\Psi_0(r)$ and $\Psi(r)$ have a maximum at certain values of $r$.
Let us estimate the average size of the school radius in each model by
calculating the positions of these maxima.
The maximum of $\Psi_0(r)$ occurs at
$R_0 = (2b^2N/3)^{1/2}$.
The maximum of $\Psi(r)$ occurs at $R$,
which satisfies the following equation obtained by differentiating the
logarithm of Eq.(\ref{eqn:prob-excluded-radius}):
\begin{equation}
 \left(\frac{R}{R_0}\right)^5
  -\left(\frac{R}{R_0}\right)^3
  =
  \frac{9\sqrt{6}}{16}\frac{w \sqrt{N}}{b^3}.
  \label{eqn:diff-log}
\end{equation}
If $N \gg 1$,
the second term on the left-hand side of
Eq.(\ref{eqn:diff-log}) can be neglected and hence
\begin{equation}
 R
  \simeq
  R_0
  \left(\frac{w\sqrt{N}}{b^3}\right)^{1/5}
  \propto
  N^{3/5}.
  \label{eqn:rouse-excluded-volume-effect}
\end{equation}
The exponent $3/5$ is the very value extracted through the data
collapse.

The scaling in the relationship between geometric dimensions and biomass
of pelagic fish schools is analogous to that developed in polymer
physics \citep{deGennes79}.
The result suggests that the dynamics are unexpectedly ``simple'' and
depend primarily on common cohesive motion in these animate and
inanimate systems.
Though the above is a very rough theory of school conformation with the
excluded volume effect,
the overall statistical properties do not depend on the details of the
model, which is a consequence of universality \citep{Stanley95}.

\subsection{Exploiting the analogy to a polymer chain model}

As indicated in the Introduction, large internal variations in packing
density occur within a school.
\citet{Freon-Misund99} pointed out that a source of substantial
variation in internal school structure is the formation of subgroups.
Such subgroups have been observed in saithe schools \citep{Partridge81},
and in schools of minnow {\it Phoxinus phoxinus} \citep{Pitcher73} and
herring \citep{Pitcher-Partridge79}.
Relatively independent movements of such clusters of individuals can
open up empty spaces and cause large variation in school volume.

The polymer-chain analogue of the subgroup in school conformation is the
``blob'' \citep{deGennes79}:
the polymer chain behaves as a series of blobs.
Based on the blob concept, for small value of the excluded volume $w$, a
subgroup within a school, with a number of
$\lambda_{\mbox{\tiny B}}$ of fish, must be nearly Gaussian.
We see this from Eq.(\ref{eqn:prob-excluded-radius}) when we find no
effect of $w$ if $w\lambda_{\mbox{\tiny B}}^{1/2}/b^3 <1$.
There is a certain value of
$\lambda_{\mbox{\tiny B}}\;[\sim (b^3/w)^2]$,
beyond which excluded volume effects become important.
A single school will appear Gaussian at scales
$r < r_{\mbox{\tiny B}}$ where
\begin{equation}
 r_{\mbox{\tiny B}}
  \simeq
  b \lambda_{\mbox{\tiny B}}^{1/2}
  \simeq
  b^4/w,
\end{equation}
while at scales $r > r_{\mbox{\tiny B}}$ it will show excluded volume
effects.
According to the ``blob'' approach, an $N$-sized school can be described
as a cluster of $N/\lambda_{\mbox{\tiny B}}$ subgroups.
Inside the subgroup
the core repulsion (by excluded volume effects) is a weak perturbation
leading to a Gaussian state.
The school radius is then written as
$R\simeq r_{\mbox{\tiny B}} (\lambda_{\mbox{\tiny B}}^{-1}N)^{\nu}$
in the native state.

Notice the plasticity of school geometry and internal structure.
The compact, dense packing, as observed in artificial environments like
small tanks, may be caused by the very strong stress.
When the school is confronted by danger such as predators, the
inter-fish distance decreases rapidly and all the vacuoles within the
school collapse quickly \citep{Freon-etal92}.
This change may be closely related to the coil-globule transition in a
polymer solution \citep{Ptitsyn-etal68,deGennes75}.

\begin{figure}[tb]
 \centering
 \includegraphics[width=6.5cm]{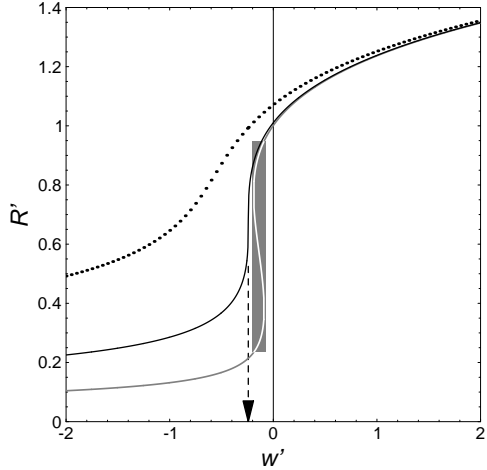}
 \caption{
 Swollen-dense packing transition in the school geometry
 [the solution of Eq.(\ref{eqn:diff-log-2})].
 Here $R'=R/R_0$ is plotted as a function of a combined variable
 $w'=(9\sqrt{6}/16)\; wN^{1/2}/b^3$.
 Shown is the transition region $|w|N^{1/2}/b^3\approx 1$.
 At $\epsilon_v = \epsilon_{\Theta}$ the excluded volume parameter $w$
 equals $0$.
 The repulsive excluded volume effect then balances the attractive
 forces between fish, and the school behaves as a Gaussian school.
 The property of the swollen-dense packing transition is quite different
 in the two regions separated by the critical value
 $w_{\vartriangle}/b^6 = 0.0228$.
 The transition between the the swollen conformation and the compact
 packing is gradual when $w_{\vartriangle}/b^6 > 0.0228$ (dotted line),
 while it becomes extremely sharp and the transition is discontinuous
 when $w_{\vartriangle}/b^6 < 0.0228$ (gray line).
 The solid line shows the marginal transition behavior, and 
 the crossover takes place when $w' = -0.242$ (indicated by the arrow).
 Notice that the solution of Eq.(\ref{eqn:diff-log-2}) is  multivalued
 [depicted by a sigmoid curve (white cutout)], showing a  first-order
 discrete phase change  \citep{Ptitsyn-etal68,deGennes75},
 when $w_{\vartriangle}/b^6 < 0.0228$.
 }
 \label{fig:4}
\end{figure}

In real schools the nature of the short range interaction is quite
complicated like van der Waals-type inter-molecular forces.
Accordingly, all fish within a school interact via the two-body
potential deformed from a parabolic potential of harmonic spring
[grouping forces have been expressed as the gradient of a potential
function in previous model studies \citep{Mogilner-etal03}].
The potential change, denoted by $\Delta U_{ij}$, is given as the
correction term to cohesive interaction potential between the $i$-th and
the $j$-th fish:
$\Delta U_{ij}$ will include steric effects and also may involve weak
attractions.
The correction factor to the probability distribution $\Psi(r)$ of the
school radius $r$, in place of Eq.(\ref{eqn:prob-no-overlap-1}), is then
given by the configuration integral
\begin{equation}
 p(r)
  \simeq
  r^{-3N}
  \int\!\!\cdots\!\!\int
  \exp \left(
	-J\sum_{i<j}
	\Delta U_{ij}/\epsilon_v
      \right)
  \mbox{d}\boldsymbol{x}_1\cdots\mbox{d}\boldsymbol{x}_N,
  \label{eqn:prob-no-overlap}
\end{equation}
where
$\boldsymbol{x}_i$ denotes the position vector of the $i$-th fish, and
$\exp (-J\Delta U_{ij}/\epsilon_v)$
is integrated over the school configuration space.
We can estimate $p(r)$ by means of the cluster expansion
\citep{Mayer-Mayer40,Uhlenbeck-Ford62}.
Introducing a function of separation between pairs,
$\boldsymbol{r}_{ij} =\boldsymbol{x}_j-\boldsymbol{x}_i$,
\begin{equation}
 \zeta_{ij}
  =
  \exp \left(-J\Delta U_{ij}/\epsilon_v \right) -1,
\end{equation}
we integrate Eq.(\ref{eqn:prob-no-overlap}) in the Mayer cluster
expansion over diagrams up to the second irreducible cluster,
yielding
\begin{equation}
 p(r)
  \simeq
  \exp\left(
       -\frac{wN^2}{2r^3}
       -\frac{w_{\vartriangle}N^3}{6r^6}
       \right)
\end{equation}
with the excluded volume parameter
\begin{equation}
 w
  =
  -\int \zeta_{12} \mbox{d}\boldsymbol{r}_{12},
  \label{eqn:second-virial}
\end{equation}
and the other parameter defined by
\begin{equation}
 w_{\vartriangle}
  =
  -\int\!\!\int
  \zeta_{12} \zeta_{13} \zeta_{23}\;
  \mbox{d}\boldsymbol{r}_{12}\mbox{d}\boldsymbol{r}_{13}.
\end{equation}
The actual configuration is decided by maximizing 
the distribution function $\Psi(r)=\Psi_0(r)p(r)$,
and this maximization condition leads to the equation determining the
school radius $R$,
\begin{equation}
 \left(\frac{R}{R_0}\right)^5
  -\left(\frac{R}{R_0}\right)^3
  -\frac{27w_{\vartriangle}}{16 b^6} \left(\frac{R}{R_0}\right)^{-3}
  =
  \frac{9\sqrt{6}}{16}\frac{w \sqrt{N}}{b^3}.
  \label{eqn:diff-log-2}
\end{equation}
Notice that, when the first irreducible cluster integral in Mayer's
expansion is only taken into account, the configuration integral,
Eq.(\ref{eqn:prob-no-overlap}), is evaluated by
Eq.(\ref{eqn:prob-no-overlap-1}).
In case that $\Delta U_{ij}$ consists of a hard core repulsion and a
short-range weak attraction
(i.e. at the distance of near-collision ($r\sim w^{1/3}$) the
attraction gradient between fish becomes steeper),
Eq.(\ref{eqn:second-virial}) is estimated as
$w=w_0 (1-\epsilon_{\Theta}/\epsilon_v)$,
expressing the dependence of $w$ on the velocity-fluctuation
$\epsilon_v$ of fish,
where $w_0$ and $\epsilon_{\Theta}$ are constants
\citep{Doi-Edwards86}.
During predation threats or laboratory observations,
the value of excluded volume parameter $w$ will change sign from
positive to negative, provided that 
the velocity fluctuation $\epsilon_v$ is suppressed below a certain
value $\epsilon_{\Theta}$.
Decreasing $\epsilon_v$ below $\epsilon_{\Theta}$,
the dimensional size of the school becomes much smaller than that of a
Gaussian school, as depicted in Fig.\ref{fig:4}.
According to Eq.(\ref{eqn:diff-log-2})
the so called ``expansion factor'' $R/R_0$ is determined not by $w$ but
by $wN^{1/2}$,  and so if $N$ is large only a small change in
$\epsilon_v$ will cause a big change in dimensional size.
For example, for school of $10^6$ fish, a small variation in
$\epsilon_v$ will induce a dramatic change in the school radius.
Equation~(\ref{eqn:diff-log-2}) states that
when $-wN^{1/2}/b^3 \gg 1$, $R/R_0$ is proportional to
$(-wN^{1/2})^{-1/3}$ and hence
the solution is as follows:
\begin{equation}
 R
  \simeq
  R_0
  \left(\frac{-wb^3\sqrt{N}}{w_{\vartriangle}}\right)^{-1/3}
  \propto
  N^{1/3},
  \label{eqn:globule-state}
\end{equation}
which shows the closest packing of fish in a school, i.e.
the constant density independent of the number of individuals.
This change may be called the swollen-dense packing transition of
schools.
%

\subsection{Configuration inside a swollen school}

The approach of treating fish schools as interacting particle systems
naturally leads to the idea of applying successful methods of
statistical physics to the description of moving together without a
leader \citep{vicsek01}.
In statistical physics, the presence of non-trivial scaling is usually
taken to mean that the dynamics are largely governed by simple geometric
properties of the system and do not depend strongly on detailed
properties of the system components \citep{wilson83}.

As for the school geometry, the basic units are the effective inter-fish
length $b$ and the number of fish, $N$.
We assemble $\lambda$ neighboring fish in a school into $N/\lambda$
groups, e.g. in the sense of the Voronoi-Dirichlet
diagram \citep{dirichlet1850,voronoi08,voronoi09}.
The length between centers of neighboring subgroups is defined by
$\lambda^{\tilde\nu} b$ with an exponent $\tilde\nu$.
After such a scale change as
\begin{equation}
 N\to \lambda^{-1}N,\quad
  b\to \lambda^{\tilde\nu}b,
  \label{eqn:scale-change}
\end{equation}
the macroscopic quantities which determine the overall properties of the
school satisfy
\begin{equation}
 g(\lambda^{-1}N,\lambda^{\tilde\nu}b)
  =
  \lambda^{\chi} g(N,b),
\end{equation}
where $\chi$ is an exponent which depends on the quantity under
consideration.
For this to hold true for arbitrary $\lambda$, the function $g(N,b)$ must take
the scaling form
\begin{equation}
 g(N,b)
  =
  N^{-\chi} g^{\mbox{\scriptsize (sc)}} (bN^{\tilde\nu}),
\end{equation}
where a new function $g^{\mbox{\scriptsize (sc)}}$ is introduced.
This is the Widom-Kadanoff scaling law in critical
phenomena \citep{Goldenfeld92}.
For example, the average radius of the school, $R$,
should be unaltered under the transformation of
Eq.(\ref{eqn:scale-change}):
\begin{equation}
 R(N,b)
  =
  R(\lambda^{-1}N,\lambda^{\tilde\nu}b)
  =
  R^{\mbox{\scriptsize (sc)}}(bN^{\tilde\nu}).
\end{equation}
Because of Eq.(\ref{eqn:scaling-school-geometry}),
$R^{\mbox{\scriptsize (sc)}} (x) = x$
and we must have
\begin{equation}
 \tilde\nu = \nu,
\end{equation}
where $\nu = 3/5$ in three dimensions.

As an example of the application of the above scaling law,
let us consider the pair correlation function inside a school,
where fish movement take place in a three-dimensional space.
A pair correlation function $\rho_2(r)$ is defined as follows.
We pick one fish at random in the school, and we place it at origin.
Then we ask, what is the number density of other fish at a (directional)
distance $\boldsymbol{r}$ from the first, and we average the result over
all choices of the first fish (and of directions).
The function $\rho_2(r)$ has an integral which is just the total number
of fish per school,
$\displaystyle{\int \rho_2(r) \mbox{d}\boldsymbol{r} = N}$.
Hence, from dimensional analysis, we can write
\begin{equation}
 \rho_2(r)
  =
  R^{-3} g(N,r/b).
\end{equation}
Under the transformation that neighboring $\lambda$ fish are grouped to
form one subunit in a school,
$\rho_2(r)$ will be reduced by $1/\lambda$, since the pair correlation
function is proportional to the number density of fish.
Therefore
\begin{equation}
 g(\lambda^{-1}N,r/\lambda^{\nu}b)
  =
  \lambda^{-1} g(N,r/b).
\end{equation}
Then, the function $\rho_2(r)$ obeys a simple scaling rule
\begin{equation}
 \rho_2(r)
  =
  N R^{-3}
  \rho_2^{\mbox{\scriptsize (sc)}}(r/R),
  \label{eqn:pair-correlation-scaling}
\end{equation}
where $\rho_2^{\mbox{\scriptsize (sc)}}$ is a dimensionless universal
function, and $NR^{-3}$ is the dimensional
factor.
Focusing on the limit $r \ll R$, we can reach the form of
$\rho_2(r)$ by a simple argument.
In a sphere of radius $r$ we have a certain number of fish,
$\tilde{N}$, related to $r$ by the excluded volume exponent:
$\tilde{N}^{3/5} b \simeq r$.
The function $\rho_2(r)$ scales like the density of fish in the
sphere,
\begin{equation}
 \rho_2(r)
  \simeq
  \frac{\tilde{N}}{r^3}
  \simeq
  \frac 1{r^{4/3} b^{5/3}}
  \quad
  (\mbox{for}\,\, r<R),
  \label{eqn:pair-correlation}
\end{equation}
which gives $\rho_2^{\mbox{\scriptsize (sc)}}(x) \simeq x^{-4/3}$.

The possible observable property,
Eq.(\ref{eqn:pair-correlation-scaling}), tells us that if we were
to measure $\rho_2(r)$ for fish schools of different size $N$,
there would be superposition of the curves obtained by plotting
$\rho_2(r)R^3 /N$ against $r/R$.
Thus this kind of scaling relation may be verified experimentally on
large schools of pelagic fish.

\section{Discussion}

The power-law scaling generally exists in dimension-to-biomass
relationship of pelagic fish schools in nature.
Here I have estimated the power-law exponent $\nu$ of the geometric
relation, based on the dependence of the power-law regime of the
school-dimension distribution $W^{\mbox{\tiny (G)}}(R)$ on $\nu$,
i.e. Eq.(\ref{eqn:depend-cutoff-nu}).
We have tested whether the distributions $W^{\mbox{\tiny (G)}}(R)$ are
self-similar (i.e. exhibit scaling), relying on both the power law for
the dimension-to-biomass relationship of
Eq.(\ref{eqn:scaling-school-geometry}) and the FSS relation in the
power-law school-biomass distribution of
Eq.(\ref{eqn:biomass-distribution}).
Plotting the scaled histogram-data from the 22 {\it in-situ}
observations, we have found that the 22 curves do indeed collapse onto
each other (Fig.\ref{fig:2}a), suggesting that $W^{\mbox{\tiny (G)}}(R)$
follows a universal functional form
[Eq.(\ref{eqn:dimension-distribution})].
We have extracted the power-law exponent of dimension-to-biomass scaling
relationship via a minimization of a measure to quantify the nature of
scaling collapse.
Next I have explained the value of the exponent $\nu$, proposing the
Gaussian school model for the fish with excluded volume $w$.
By exploiting the analogy between fish-school and polymer
conformations, we have examined the behavioral algorithm governing the
swollen conformation of large-sized schools.
We have seen that the exponent is modified strongly by the steric
effect.
What is universal in Eq.(\ref{eqn:scaling-school-geometry}) is the
exponent $\nu\approx 0.6$:
it is independent of species as well as environmental conditions, and
the same for all schools.
The constant that multiplies $N^{\nu}$ in
Eq.(\ref{eqn:scaling-school-geometry}) is non-universal
[$\sim (wb^2)^{1/5}$ with effective inter-fish distance $b$, predicted
from Eq.(\ref{eqn:rouse-excluded-volume-effect})],
and depends on the details of interactions between fish which may vary
with species and environmental conditions \citep{Morgan88}.

To understand the scaling law for school geometry,
Eq.(\ref{eqn:scaling-school-geometry}) with $\nu =3/5$,
it is essential to see what the value of $\nu$ represents.
Obviously, for a regular object embedded into a $d$-dimensional
Euclidean space, Eq.(\ref{eqn:scaling-school-geometry}) would have the
form $N(R)\sim R^d$ expressing the fact that the volume of a
$d$-dimensional object grows with its linear size $R$ as $R^d$.
Contrary to an integer dimensionality, it has been shown that the
packing structure within the schools is characterized by a non-integer
(i.e. fractal) dimensionality of $1/\nu \approx 1.7$:
the number of fish in a school of radius $R$ scales as
$N(R)\sim R^{1/\nu}$.
During the last decades of the twentieth century
it has widely been recognized by researchers working in diverse areas of
science that many of the structures commonly observed possess a rather
special kind of geometrical complexity;
the name ``fractal'' was coined by Benoit Mandelbrot
for these complex shapes \citep{Mandelbrot82}.
Objects of biological origin are many times fractal-like
\citep{vicsek01-book}.

FSS plots of data lead to possibly important and
practical insights in ecology.
In the context of the fisheries mission,
the demonstrated geometric relation between dimensions and biomass of
pelagic fish schools, when applied to mapping pelagic schooling fish,
will largely improve the precision in the fish stock assessment.
Because of the linear dependence of the fish population on
$\langle N \rangle_P$,
as given by Eq.(\ref{eqn:cutoff-population}),
the fish stock abundance can be inferred from an index
(i.e. cut-off size) that can be determined directly from observations
[see Eq.(\ref{eqn:cut-off-size})].

I am very grateful to Leah Edelstein-Keshet for helpful suggestions.


%
\end{document}